\documentclass[letterpaper,USenglish]{lipics-v2021}

\nolinenumbers
\hideLIPIcs  

\usepackage{url} 
\hypersetup{colorlinks=true,linkcolor=red,urlcolor=blue,citecolor=blue}

\usepackage{tabularx} 
\usepackage{booktabs} 

\usepackage[maxfloats=150]{morefloats}                                               
\usepackage{microtype}                                                               
\usepackage[utf8]{inputenc}   
\usepackage{courier} 

\usepackage{graphicx} 
\usepackage{amssymb} 
\usepackage{txfonts} 

\newcommand\Invisible[1]{                                                            
  \marginpar{\color{white}{\fontsize{.5}{.5}\selectfont #1 }}                        
}  

\newcommand{\Exclude}[1]{}

\newcommand\BoldSection[1]{\vspace{0.5 \baselineskip} \noindent \textbf{#1} \noindent}
\definecolor{Gray95}{gray}{0.95}
\definecolor{forestgreen}{rgb}{0.13, 0.55, 0.13}

\newcommand{\AtFoot}[1]{\let\thefootnote\relax\footnotetext{{#1}}} 

\newcommand{\remove}[1] {}

\usepackage{microtype}   %if unwanted, comment out or use option "draft"
\usepackage{listings} 
\usepackage[scaled]{beramono} 
\usepackage{bera} 
\usepackage[T1]{fontenc}
\usepackage{changepage} 

\usepackage[iso,american,cleanlook]{isodate}                                         
\usepackage[export]{adjustbox} 
\usepackage{fancyhdr}                                                             
\makeatletter\let\ps@plain\ps@fancy\makeatother
\usepackage{background}
\backgroundsetup{
  angle=0,  
  scale=1, 
  opacity=1,
  firstpage=true,
  position=current page.south west, 
  hshift=84pt,
  vshift=460pt,
  contents={\colorbox[rgb]{0.99,0.78,0.07}{Page \thepage}}
}

\makeatletter
\lst@Key{countblanklines}{true}[t]%
    {\lstKV@SetIf{#1}\lst@ifcountblanklines}

\lst@AddToHook{OnEmptyLine}{%
    \lst@ifnumberblanklines\else%
       \lst@ifcountblanklines\else%
         \advance\c@lstnumber-\@ne\relax%
       \fi%
    \fi}
\makeatother

\lstdefinestyle{numbers}
{numbers=left, stepnumber=1, numberstyle=\tiny, numbersep=10pt}
\lstdefinestyle{nonumbers}
{numbers=none}

\usepackage{tikz} 
\usepackage[inline]{enumitem}

\usepackage{bm} 

\usepackage[sc]{mathpazo} 

\usepackage[ampersand]{easylist}

\usepackage[margin=15pt,font=small,labelfont={bf,sf}]{caption}

\title{Exploring Time-Space trade-offs for \emph{synchronized} in Lilliput} 

\author{Dave Dice}{Oracle Labs}{dave.dice@oracle.com}{https://orcid.org/0000-0001-9164-7747}{}
\author{Alex Kogan}{Oracle Labs}{alex.kogan@oracle.com}{https://orcid.org/0000-0002-4419-4340}{}
\authorrunning{D. Dice and A. Kogan} 

\Copyright{Oracle and/or its affiliates}


\category{}%optional, e.g. invited paper

\supplement{}%optional, e.g. related research data, source code, ... hosted on a repository like zenodo, figshare, GitHub, ...

\funding{}%optional, to capture a funding statement, which applies to all authors. Please enter author specific funding statements as fifth argument of the \author macro.

\EventShortTitle{}

\begin{document}

\maketitle

\begin{abstract}
In the context of Project \emph{Lilliput}, 
which attempts to reduce the size of object header in the HotSpot Java Virtual Machine (JVM), 
we explore 
%% the space-time properties of 
a curated set of synchronization algorithms.  Each of the algorithms could serve as 
a potential replacement implementation for the ``synchronized'' construct in HotSpot. 
Collectively, the algorithms illuminate trade-offs in space-time properties.  
%% aspects of the space-time trade-off.  

The key design decisions are \emph{where} to locate synchronization metadata (monitor fields), 
\emph{how} to map from an object to those fields, and the lifecycle of the monitor information.  

The reader is assumed to be familiar with current HotSpot implementation of ``synchronized'' 
as well as the Compact Java Monitors (CJM) design \cite{CJM} and Project Lilliput \cite{Lilliput}.  
\end{abstract}

\Invisible{Oracle Invention Disclosure Accession : ORC2133570-US-PSP} 
\Invisible{Explore design space} 
\Invisible{Selected; Curated; } 

%% Would strongly prefer to use proper header/footers ..
%% \AtFoot{\today \hspace{1mm} $\blacklozenge$ \hspace{1mm} Copyright Oracle and or its affiliates} 
%% \AtFoot{\today \hspace{1mm} \textbullet \hspace{1mm} Copyright Oracle and or its affiliates} 

%% \section{Introduction} 

\section{Algorithms}

All the candidate locking algorithms below are FIFO/FCFS-fair, unless stated otherwise,
 and are able to support the full gamut of
\texttt{synchronized} operators.  All are space-conserving in the sense that 
acquiring a monitor requires that a thread contribute a \emph{lock record} to some
set of records associated with the object, and releasing that monitor reclaims that same
lock record, avoiding the \emph{objectMonitor} accretion concerns attendant in the existing
HotSpot monitor implementation. 

\subsection{Candidate : HashChains}
In this variant, \texttt{synchronized} operations do \underline{not} access the object header.  
The header word is never displaced by synchronization and no bits in the header
are reserved for synchronization. 
This approach is appealing as avoids it multiplexing and overloading of the header word,
simplifying the encoding thereof and reducing couplings between the synchronization subsystem
and other components on the JVM.  

%% No information in the header word is displaced by synchronization. 

\Invisible{unfettered by}  

The JVM maintains a shared hash table of synchronization buckets.  Each bucket
contains a simple mutex lock, and a pointer to the head of a linked list of \emph{lock records}.  
Similar constructs in the linux kernel -- \emph{futex} wait channels -- are sized at startup time, based on the number of 
logical processors.  For all experiments reported herein, we used 4096 buckets.  
If necessary the hash tables could be resized at runtime.  

To acquire a monitor, a thread first allocates and constructs a lock record, and then
identifies the bucket associated with the object.  We can hash the virtual address of the object
to map to the bucket, although this would require rehashing in the event of moving garbage collections.
To avoid that, we could instead hash on the object's identity hashCode value, although this
would necessitate assigning identity hashCode values on objects that participate in synchronization. 
Our native C++ implementation hashes on the virtual address.
The thread next acquires the bucket lock, and emplaces its lock record on the chain.  
\Invisible{Forms, Constitutes, represent, encode} 
The bucket locks can be simple \texttt{pthread} mutex locks, or any other simple native lock, 
such as MCS locks \cite{tocs91-MellorCrummey}. 
Our implementation uses \emph{hemlock}\cite{Hemlock-arxiv,Hemlock-SPAA}.  
The thread then determines if there are any conflicting lock records on the chain 
-- lock records inserted by other threads that refer to the same object.
Subsequently, the thread releases the bucket lock.  
If no conflicting elements were observed, then the thread acquired the monitor 
without contention and is the
owner of the monitor, and as such, may enter the critical section without waiting. 
Otherwise the thread parks, waiting on a field in its lock record to change state indicating
that it has been granted ownership.  

The lock record posted by a thread continues to reside on the chain while the thread executes
in the critical section.  There is no singular central memory location that encodes if an object is locked,
but rather the \emph{locked-ness} is determined by the presence or absence of conflicting
lock records on the object's hash chain.  Lock records include a ``state'' field which indicates if the 
associated thread holds the lock, is waiting on the lock, or is in \texttt{wait()}.  
The set of lock records resident in the bucket and associated with a given object form
the \emph{entry set} for that object. 

To release a monitor, a thread again acquires the bucket lock and removes the element
it originally posted, moving that element to a thread-local free list, allowing for subsequent re-use.
As threads typically hold a very small number of locks concurrently, we can use 
thread-local free lists (stacks) of free lock records. 
The thread also checks for and identifies a successor from the set of lock records
associated with the object.  If no successor is found, the thread simply
drops the bucket lock, otherwise it marks the successor's lock record as being the 
current owner, releases the bucket lock,  and unparks the successor.

To allow for efficient \texttt{IllegalMontorStateExceptions} checks, and for nested locking, and automatic 
unlocking of monitors, each thread maintains a list of lock records reflecting the monitors
it currently holds.  

In our implementation, instead of a simple unstructured ``bag'' of lock records on the bucket chain,
we use a \emph{spine-and-rib} design where the spine elements reflect the current owner,
and all remaining threads waiting on that object are linked as ribs off that spine element, 
in order to reduce traversal times and thus reduce hold times for the bucket locks.  

Lock records could be implemented as native C++ constructs, or as first-class Java objects.
In the case of the latter, much of the synchronization subsystem could be shifted into pure Java code.

No type-stable memory (TSM) or safe memory reclamation (SMR) is required.  
Furthermore the design places tight bounds on the amount of memory required for synchronization.   
No explicit deflation step is required.  Trimming of thread-local free lists of lock records,
if it is every needed -- which is unlikely -- is a strictly thread-local decision.  

While simple, this approach entails a number of performance challenges.  To acquire and release an 
uncontended monitor, we need to acquire and release the bucket lock twice, once to post the
lock record to the chain, and another to extract it.  This results in poor latency compared to other approaches.
The traffic on the chain locks arising from such  ``double locking" also impinges on scalability.  
(Locks that are implemented with an ``inner'' or ``meta'' lock to protect their queues are
usually inferior in performance locks that avoid inner locks).  
In addition, we're exposed and vulnerable to \emph{false contention} on the bucket locks because of hash collisions.
Even absent bucket lock collisions, we can incur false sharing and costly coherence traffic
when multiple unrelated synchronization operations interleave accesses on a given bucket.  
Accordingly, both latency and scalability suffer.  

\Invisible{Delegation of concerns; seperation of concerns;}

\subsection{Candidate : HashChains+3}  

\textbf{HashChains+3} builds on \textbf{HashChains} but also requires 3 bits to be reserved 
for synchronization in the object's header word.
The bits encode \emph{Locked, WaitersExist, and Impatient} indicators. 
Our implementation does not currently use \emph{Impatient}, but we reserve the bit to allow 
Fissile Locks \cite{dice2020fissile,fissile-netys} to provide bounded bypass, if desired,
to improve contended performance.  
As expected \emph{Locked} indicates the lock is held.  
\emph{WaitersExist} is a conservative indication -- false-positives are allowed but
never false-negatives -- that contending threads exist on the associated hash chain. 
The indicator allows fast-path uncontended unlock operations when there are no waiting
threads.  

Arriving threads first attempt to acquire the lock by using CAS to toggle the \emph{Locked} bit
from 0 to 1.  Unlike \textbf{HashChains}, the owner's lock record does not reside on the chain.
Failing to acquire via that fast-path, threads ensure the \emph{WaitersExist} bit is set and 
emplace on the chain, under the bucket lock, and then park in the usual fashion,
waiting on the ``state'' field in its own lock record. 
The chain contains only waiting threads, and never the owner.  

To release the monitor the thread first consults the \emph{WaitersExist} bit.  
If clear, the thread attempts to use CAS to clear the lock bit, while ensuring the
\emph{WaitersExists} bit remains clear. 
If the CAS was successful, no further actions are required.  
Otherwise, if \emph{WaitersExists} was set, the thread acquires the bucket lock and
detaches a successor, if any.  If no additional conflicting threads (beyond the successor)
are present on the chain, it also clears the \emph{WaitersExist} bit.  The thread 
then drops the bucket lock, marks the successor's lock record as being the new owner, 
and unparks the successor, passing ownership directly to the successor. 
The \emph{Locked} bit remains held continuously while ownership is conveyed.  
If no successor was found on the chain, the thread clears both the \emph{Locked} and 
\emph{WaiterExist} bits and drops the bucket lock.  

Critically, uncontended acquire and release operations do not need to access the hash chains,
improving latency.  The hash chains are needed only under contention.  
Related ideas worthy of note can be found in \texttt{WebKit} \cite{webkit-locking}. 

\subsection{Candidate : CJM} 

\textbf{Compact Java Monitors (CJM)} \cite{CJM} are based on the Compact NUMA-Aware Locks (CNA) algorithm, 
but forego the \emph{NUMA-Aware} property and focus instead on the \emph{Compact} aspect.  CNA is itself a 
variation on the gold-standard MCS (\underline{M}ellor-\underline{C}rummey \underline{S}cott) 
\cite{tocs91-MellorCrummey} queue-based lock algorithm 
\footnote{If you're unfamiliar with CJM, please see \cite{CJM}. That document
also includes a description of the MCS algorithm in an appendix. Briefly,
MCS uses atomic \texttt{SWAP} and \texttt{CAS} primitives to construct a lock-free queue --
implemented as a singly linked list -- of elements, where each element serves to
represent an allocation request by the thread that posted that element.  
The head element is the current owner and the MCS lock word points to the tail.}. 
%% From page one of \url{https://arxiv.org/pdf/2003.05025.pdf}   

%% \symbol{"200B} 

\Invisible{Underlying much of the following design is our approach from Compact NUMA-Aware Locks.} 
CNA was published in EuroSys 2019 \cite{EuroSys19-CNA,arxiv-CNA}  and has been implemented
in both user-space in in the Linux kernel as a replacement for the existing low-level
%% being integrated into the Linux kernel as a replacement for the existing low-level
\emph{qspinlock} construct \cite{linux-locks,Long13}, which is itself based on MCS.  
\Invisible{CNA is itself a variation on classic MCS.}  
One of the key ideas in 
CNA is propagating values of interest from the MCS owner's queue element into the successor,
which allows the lock body to remain compact -- just one word.  
Specifically, fields that would normally appear in the body of a lock are
instead maintained in the owner's queue element and, at unlock-time, conveyed to the successor in the queue. 

In the context of the current discussion we're not interested in NUMA-aware aspects of CNA, where we 
propagate the list of remote queue elements 
%% isolated on a distinct chain,  
through the MCS chain (avoiding extra fields in the lock body), but instead
leverage the fact that the lock body is compact.  Taken to the extreme, CJM shifts \emph{all}
the fields that would normally reside in the classic HotSpot \emph{objectMonitor} construct 
into the MCS queue elements, so we can represent the abstract Java monitor with just a single pointer 
to the MCS tail.

When an object is locked, the implementation relocates the 
entire header word into a \emph{displaced header} which is conveyed through the chain.
The identity hashCode value will be assigned
on the first synchronization operation on a given instance, in order to avoid mutating the
displaced header.  Relatedly, we assume the encoded class (\emph{klass}) information in the header
is immutable, or at least rarely mutated.  And finally, we assume that the GC \emph{age} bit field
in the header word is also mutated infrequently
\footnote{The object header fields typify Conway's Law \cite{conways-law}.}.  
See Section 6 of \cite{CJM} for details.

Accessing the header fields (identity hashCode value, class information, etc) is relatively 
simple under CJM.  First, if the object is not engaged in synchronization, those fields
reside in their usual ``home'' position in the object header.  If the object happens to be
locked by the thread attempting to fetch the header, which is easily determined, the caller can
quickly extract the displaced header value from the queue element (lock record) it originally 
posted to the MCS chain.  
Finally, if the object is locked by some other thread, which we expect to be rare, we can 
use the access protocol described in \cite{CJM} Section 6 to extract the header word value. 
This final mode uses a chase-and-capture idiom that enjoys obstruction-free progress properties.  

Briefly, CJM provides the following desirable properties:
\begin{itemize}
\item[$-$] Eliminates stack-locking, resulting in a simple unified encoding in the header, and only one
flavor of locking.  
\item[$-$] Performance on-par with the existing subsystem.
\item[$-$] Extremely simple with fewer lines of code, and easily maintainable.  
\item[$-$] FIFO-fair admission policy, whereas the existing system admits unbounded bypass and starvation. 
\item[$-$] Reduced interactions and dependencies on other subsystems : safepoint, GC, etc. 
\Invisible{Decoupled} 
\item[$-$] Avoids accretion of objectMonitors and deferred deflation to recover of those
monitors.  Threads contribute
one queue element to a monitor's queue when they acquire the lock, and recover that same element
when they release the lock, resulting in space-conserving \emph{self-cleaning} pay-as-you-go memory use with
extremely tight bounds.  This property is a key desideratum and improvement over the existing 
\texttt{synchronized}  implementation.  
\item[$-$] As described, CJM is FIFO, but to gain more performance we can allow bounded bypass
using the \emph{Fissile Locks} technique  \cite{dice2020fissile,fissile-netys}.  
\end{itemize} 

\Invisible{CJM : All avoid inflation and need for STW deflation;  deflate agro ASAP;} 
\Invisible{Reduced complexity} 

\section{Waiting Policies} 

The choice of waiting policy -- how a thread waits to acquire a contented lock -- 
has subtle ramifications.  

First, when executing in native C++ code we need to respect HotSpot's ``no loitering'' invariant.
Threads must not busy-wait indefinitely.  Bounded spinning is acceptable, but threads waiting
for some condition to be satisfied must be prepared to resort to parking after some period.  
In addition, indefinite polling via operating system provided primitives such as
\texttt{sched\_yield} is not sufficient.  Polling via timed sleeping is also not acceptable,
as operating system timers don't scale and can also prevent systems from entering low power states. 

In pure Java code, indefinite spinning is permissible but unwise.  Say, for instance,
that a lock implemented in pure Java spins indefinitely without parking.  Furthermore
say that we have more ready threads than logical CPUs, so operating system preemption
(involuntary preemption) is in play.  If the lock holder is preempted then we have to 
wait for spinning threads to be preempted and the CPU to be passed to the lock holder to 
achieve succession and progress.  Similar stalls can occur if the lock holder transfers ownership
to a preempted successor.   Involuntary preemption (time slicing) operates on a relatively long time frame 
(1 to 10 msecs, which is effectively geologic time) and as such, throughput over the contented 
lock will suffer.  

The current \texttt{synchronized}  implementation uses adaptive spinning where we try to use 
the success or failure rate of recent spin attempts on a given object to better inform the spin-park decision 
and spin duration. 
\texttt{ReentrantLock} does not spin and immediately parks on contention.  

Our CJM implementation uses a simple spin-then-park \underline{STP} waiting policy.  
Simple STP using a bounded spin duration of half the voluntary context switch ``round trip'' latency
is \emph{2-competitive} versus the ideal. 
That is, STP with a fixed spin duration of half the context switch latency 
is never more than $2x$ worse (performance) than an idealized schedule where we have 
an ``oracle'' with perfect foreknowledge -- a so-called \emph{offline} algorithm --  
that tells us in advance the waiting time, and thus lets us decide immediately and 
with perfect certainty whether to spin or park when a thread needs to wait for a lock.
As we don't have such foreknowledge, the spin-versus-park decision is considered
an \emph{online} problem 
\footnote{The spin-park problem is equivalent to the \emph{Ski Rental Problem} \cite{SkiRental}.}.  
The spin phase uses the PAUSE instruction on x86, although MONITOR-MWAIT would be a better
choice if available.  (PAUSE loops executed in virtual machines trigger
VM exits, which inform the VM that busy-waiting is taking place, and inform the
VM that gang coscheduling may be called for).  

Parking, instead of unbounded spinning, also reduces the number of ready threads,
and acts to forestall the onset of preemption.  

\Invisible{keywords: competitive analysis; competitive ratio; ski-rental problem;
optimal stopping; worst case analysis; online algorithm; offline algorithm; } 

\Invisible{Satisfied; condition-of-interest;} 
 
The CJM and HashChain implementations reported herein used a maximum (bounded) spin duration
of 2500 PAUSE instructions before they devolve to parking, rechecking the
condition of interest after every PAUSE instruction. 
Specifically, we execute a loop where each iteration checks (polls) the condition
of interest.  If satisfied, we return immediately.  Otherwise we execute 
a single PAUSE instruction.  After 2500 iterations if the condition is not
satisfied, we revert to parking the waiting thread. 
Spinning is used solely as an optimistic ``bet'' to avoid parking and the overheads
of context switching.  (Unfortunately the typical latency of a PAUSE instruction
varies by a factor of 10 between steppings on Intel CPUs). 
All spinning in our implementations is \emph{local} with at most one thread busy-waiting on a given
location at any given time in order to reduce coherence traffic.
In contrast, simple test-and-set and ticket locks use \emph{global} spinning. 

Busy-waiting or spinning should be performed in a \emph{polite} fashion, which minimizes
the performance impact of spinning on other threads in the system.  
Impolite naive spinning can impede the performance of other coscheduled threads.  
Crucially, we don't want the spinning thread to compete for resources such as power and thermal caps, 
turbo-mode enablement, 
cache residency, interconnect or memory bandwidth, pipelines, etc.  
On x86, for instance, the PAUSE instruction can be used for polite spinning.  
Altruistic spinning may even benefit the spinning
thread, as the lock owner may execute its critical section and relinquish the lock
in a more timely fashion, avoiding the so-called \emph{tragedy of the commons}.  
Note that we intentionally avoid using \texttt{sched\_yield} on linux
as, with the advent of per-CPU dispatch queues,  it is advisory and has no particular semantics. 
Yielding is also not polite to sibling threads co-executing on the system.  

\Invisible{Altruism; community awareness; sharing; tragedy of the commons}  

\Invisible{PAUSE and \texttt{sched\_yield} do not suffice for long-term indefinite waiting ; must park
Avoid depending on OS preemption, which operates over long time scales, for progress.} 

The \textbf{succession policy} is convolved with waiting policies.
Succession describes how a lock algorithm conveys ownership to waiting successors. 

\BoldSection{Direct handoff} transfers ownership directly from the outgoing thread
to the designated successor.  Examples include CJM, MCS, CNA, the linux kernel's 
\texttt{qspinlock} and Java's \texttt{ReentrantLock} in fair (FIFO) mode.  
All FIFO/FCFS lock algorithms use direct handoff but direct handover locks
are not necessarily FIFO.  Direct handover perform poorly when preemption
(involuntary context switching -- oversubscription) is in play, as would be the case when the number 
of ready participating threads exceeds the number of logical CPUs.  
In particular, direct handoff allows ownership to be transferred to preempted thread.  
Direct succession also performs poorly when parking is in use as the
overheads incumbent in voluntary context switching are subsumed into the critical section,
greatly increasing the effective length of critical sections.   
If we unpark a successor, it typically takes more than 10000 cycles for the wakee 
to resume and return from park (and even more if the wakee is dispatched onto 
a CPU that was previously idle in deep sleep states).  
Using a spin-then-park waiting policy can provide some relief against that problem. 
Unfortunately, if we use a spin-then-park waiting policy with a FIFO admission 
policy, the immediate successor will be the thread that has waiting longest, and is
thus most likely to have consumed its spin allotment and resorted to parking.
This can result in a rather abrupt drop in performance in certain areas of the 
parameter space, violating the \emph{principle of least surprise}. 

All the \texttt{HashChains} forms are ostensibly FIFO.  But, as they use
an additional internal lock to protect their queues, it is possible that thread 
$T1$ arrives before $T2$ in when calling $Lock(L)$ but $T2$ beats $T1$ to the
inner lock and enqueues before $T1$.  In practice this is not a significant
concern.

\BoldSection{Competitive handoff} releases ownership and, if necessary, 
``pokes'' a potential successor -- the heir apparent -- to retry the lock and ensure progress.
Competitive handoff is also known as \emph{barging, bypass, pouncing} or \emph{renouncement}.  
After the owner releases ownership in \texttt{unlock()},  newly arriving threads in 
\texttt{lock()}  can pounce on the lock, and bypass (barge past) other threads that have waited longer.   
Example of algorithms and implementations that use competitive handoff include 
Java's \texttt{synchronized},  \texttt{ReentrantLock}, \texttt{pthread\_mutex} primitives, 
test-and-set locks, etc.   Competitive handoff admits long-term unfairness and starvation.
In addition, the degree of unfairness is influenced by the fairness of the platform's
cache arbitration policy, and thus varies.   Perversely, simple 
test-and-set locks are NUMA-friendly on x86 platforms, as threads on the same
node are more likely to acquire ownership of a just released lock.  

Competitive succession, however, tends to be more tolerant of preemption,
as the successor, by definition, was on a CPU when it acted to acquire the lock,
avoiding the scenario where we pass ownership to a preempted thread as can
manifest with direct handoff.  

As a general observation, locks that use competitive handover will yield 
better aggregate throughput than those that use direct handover.  This drives
the use of competitive handover for common locks such as \texttt{ReentrantLock},
default \texttt{pthread\_mutexes} and Java's \texttt{synchronized}.   Competitive
succession, however, allows unbounded unfairness.  

Approaches such as \emph{Concurrency Restriction} \cite{arxiv-GCR, EuroPar19-GCR} 
and \emph{Malthusian Locks} \cite{eurosys17-dice,arxiv-Malthusian} act to intentionally
reduce the number of threads circulating over a lock in some period, and improving
the performance of various lock algorithms, particularly those that use direct handoff. 

We can capture the best attributes of competitive handoff and direct handoff
with hybrid schemes such as Fissile Locks \cite{dice2020fissile,fissile-netys}, which
allow competitive succession over the short term, but if waiting threads become
\emph{impatient} we switch briefly to direct handover to avoid starvation.
The aggressiveness with which waiting threads determine they have become impatient
acts as a tunable ``knob'' to strike an explicit trade-off between throughput and fairness, and
reflects the inherent tension between those properties. 
(Infinite patience is equivalent to full competitive succession, and \emph{no} patience
makes Fissile locks equivalent to directed handoff).  
The tactic provides \emph{bounded bypass}.  CJM-By (below) employs this approach. The implementation
used in this paper inhibits bypass after the longest waiting thread has waited more than
1 millisecond. 

Addition discussion on the topic of waiting and succession policies can be found in 
Section 5.1 of \cite{arxiv-Malthusian}.

\section{Empirical Results} 

\Invisible{Evaluation; SuT} 

Unless otherwise noted, all data was collected on an Oracle X5-2 system.
The system has 2 sockets, each populated with
an Intel Xeon E5-2699 v3 CPU running at 2.30GHz.  Each socket has 18 cores, and each core is 2-way
hyperthreaded, yielding 72 logical CPUs in total (2x18x2).  The system was running Ubuntu 20.04 with a stock
Linux version 5.4 kernel, and all software was compiled using the provided GCC version 9.3 toolchain
at optimization level ``-O3''.
64-bit C++ code was used for all experiments.
Factory-provided system defaults were used in all cases, and Turbo mode \cite{turbo} was left enabled.
In all cases default free-range unbound threads were used (no pinning of threads to processors).

All the underlying native C++ locking algorithms support the full gamut of synchronization 
and hashCode operations. 
This framework -- the locks, exposing a common API, and the associated benchmarks -- 
serve as a useful and faithful in-vitro model for the performance of \texttt{synchronized} 
activities in HotSpot.  The locking algorithms are implemented via portable C++ \texttt{std::atomic} 
primitives and a park-unpark interface based on per-thread mutex-condvar pairs.  
By using \texttt{std::atomic}, no explicit memory fences or barriers were required. 
To reduce false sharing, each lock record was sized at 128 bytes and aligned accordingly.

We collected performance data with a synthetic \texttt{MuteBench}  micro-benchmark, which
measures performance under lock contention.   
When we transliterate the benchmarks from C++ to Java, we see that the C++ CJM algorithms yields
about the same performance as the existing HotSpot \texttt{synchronized}  implementation. 
We note that the existing HotSpot \texttt{synchronized}  implementation is superior to equivalent
\texttt{pthreads} code for both contended and uncontended operations, a property expected by
developers and which we want to retain.  
\footnote{Say we have group of threads contending on a single high-traffic \texttt{pthread} mutex, where
threads arrive and depart frequently.  
The user-mode \texttt{pthread} mutex implementation in \texttt{glibc} does not use spinning,
so contended threads immediately resort to blocking in the kernel via the \emph{futex} mechanism.
The linux kernel spin lock that protects the
futex hash chain associated with that user-mode mutex address will itself become highly contended, 
often to the point where most of the waiting time is for that spinlock, instead of ``normal'' waiting
for a wakeup notification.  When we implemented Compact NUMA-Aware locks (see section 7 of \cite{arxiv-CNA}) 
we observed that improved kernel spin locks (those protecting the futex chains) sped up the futex 
operations and in turn sped up apps with highly contended pthread mutexes.  
Equivalent contended \texttt{synchronized} code running in the HotSpot JVM saw no such benefit as each thread already 
blocked on its own parking construct, and therefore we avoided hot futex chains.   
Put differently, the approach used by the JVM where each thread has private parking constructs
serves to diffuse accesses over the set of kernel futex hash chains, avoiding hotspots and
conferring an advantage. 
The decay in scalability exhibited by contended pthreads mutex operations in 
Figure \ref{Figure:graph-Maximum-Contention}, below, arises from this secondary contention.}.

\Invisible{Comparable; equilibrate;  Track closely; proven faithful; in-vivo vs in-vitro} 
\Invisible{Retain; preserve; protect;} 

Note that ideal scalability, even absent and communication or contention, 
is elusive because of conflicts for shared resources, caps on energy, and thermal constraints.  
See Section 7 \emph{Maximum Ideal Scalability} of \cite{arxiv-TWA}.

The \texttt{MutexBench} benchmark spawns $T$ concurrent threads.  
We have a global pool of $NL$ shared locks. $NL$ is configured as 1 in all runs reported below.
Threads iterate as follows.  
Each thread selects $NA$ random locks, without replacement, from the set of $NL$, forming a lockset.   
(The locks are shared, but the locksets are thread-private).  
To avoid deadlock, we sort the lockset by address.   We also configure $NA$ as 1 for
all runs reported below \footnote{We took data over a wide range of $NA$ and $NL$ values
but opted to report on runs where both values were set to 1.}. 
Each thread acquires all the elements of its lockset, and then executes a critical
section of $CSL$ steps of a shared  C++ \texttt{std::mt1993} pseudo-random number generator.  
The thread then immediately releases the lockset elements in reverse order,
mimicking Java's usual last-acquired-first-released pattern.  
The thread then executes a non-critical phase where it first computes a uniform random 
value in the range $[0,NCSL*2)$ and then executes that many steps of a a thread-local
\texttt{std::mt1993} random number generator. 
The average and median duration of the non-critical phase is $NCSL$ steps of
the thread-local random number generator.  
We intentionally randomize the non-critical section duration to avoid \emph{entrainment} where threads 
would otherwise tend to enter the critical section in a relatively stable cyclic order,
which can persist for long periods, and where the number of NUMA node transitions inherent in 
that schedule  is determinative of performance.  
$NCSL$ and $CSL$ are specified as command-line arguments.
At the end of a 10 second measurement interval the benchmark
reports the total number of aggregate iterations completed by all the threads.
We report the median of 7 independent runs. 
(To assist in detecting potential safety ``exclusion'' errors, after the measurement
interval, the benchmark resets the shared PRNG state and then, in single-threaded execution, 
advances the PRNG by the total number of iterations completed, and then ensures the final 
state agrees with the state of the PRNG at the end of the measurement interval).

In the figures below the $X$-axis reflects the number of concurrently executing threads contending for the
lock, and the $Y$-axis reports aggregate throughput.
\Invisible{the tally of all loops executed by all the threads in the measurement interval}
For clarity and to convey the maximum amount of information to allow a comparison of the algorithms,
the $X$-axis is offset to the minimum score and the $Y$-axis is logarithmic.

To facilitate comparison, we also include performance data on a version of the 
\texttt{MutexBench}  benchmark 
transliterated to java, using \texttt{synchronized}, \texttt{ReentrantLock}, and
\texttt{ReentrantLock} in FIFO mode.   We used JDK version 16.0.2 with the following command-line
arguments \texttt{-server -XX:-UseBiasedLocking -Xmx10G -Xms10G  -XX:+EnableContended}.  
%% -server -XX:-UseBiasedLocking -Xmx10G -Xms10G -XX:-RestrictContended -XX:ContendedPaddingWidth=128 -XX:+EnableContended
As we also collected data with \texttt{ReentrantLock}, we sized the heap at 10Gb as that locking
algorithms allocates a new queue element for each acquisition \cite{lea-aqs} and puts
pressure on the allocator and collector.  
We employed \texttt{@Contended} to avoid false sharing between \texttt{ReentrantLock} instances and
other data elements.  
To deal with variance, each benchmark process
executes 7 sub-runs of 10 seconds each.  An external script runs the benchmark 3 times in sequence,
with completely independent processes, for total of 21 runs of 10 seconds each. 
We report the median value of those 21 runs for aggregate throughput.  

To allow comparison against simple \texttt{pthreads} we also included a form 
of \texttt{MutexBench}  where
each object contained an full embedded \texttt{pthread\_mutex} instance.

\begin{center} 

%% \begin{figure}[h!]
%% \includegraphics[width=13cm,frame]{./graph.pdf}
\includegraphics[width=13cm]{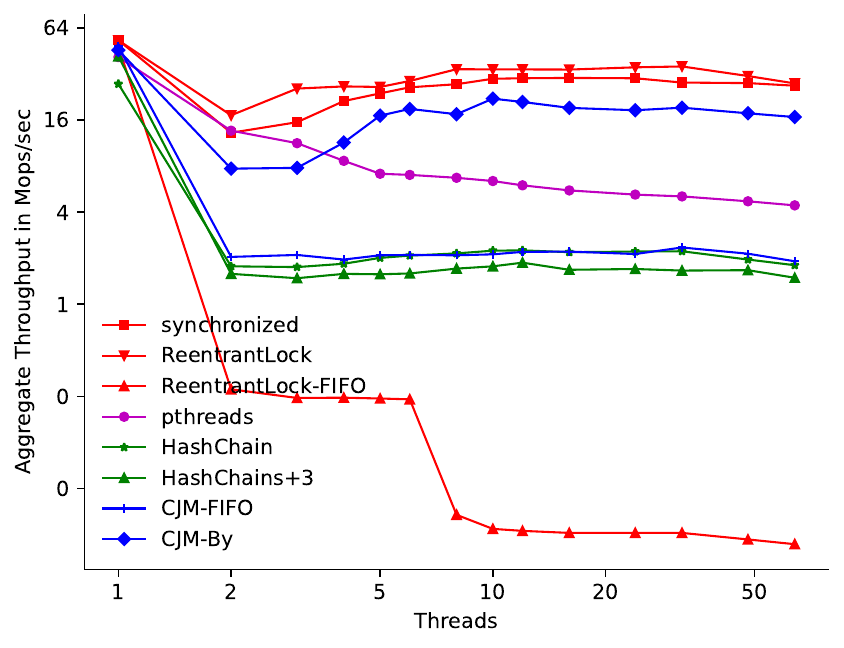} 
\vspace{-8pt}      %% reduce whitespace between graph and caption
\captionof{figure}{Maximum Contention}
\label{Figure:graph-Maximum-Contention}
%% \end{figure}

%% \begin{figure}[h!]
%% \includegraphics[width=13cm,frame]{./graph.pdf}
\includegraphics[width=13cm]{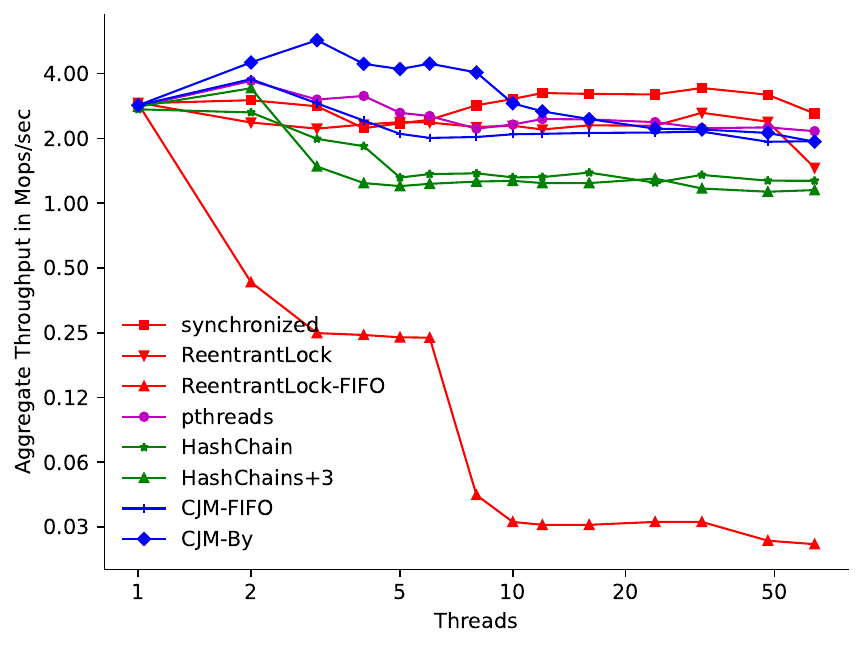}  
\vspace{-8pt}      %% reduce whitespace between graph and caption
\captionof{figure}{Moderate Contention} 
\label{Figure:graph-Moderate-Contention}
%% \end{figure}

%% \begin{figure}[h!]
%% \includegraphics[width=13cm,frame]{./graph.pdf}
\includegraphics[width=13cm]{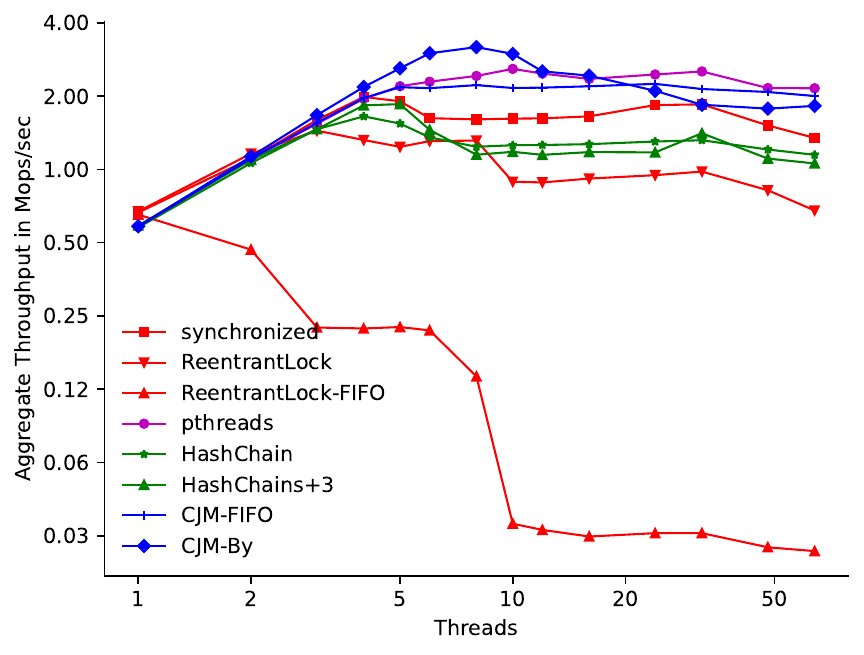} 
\vspace{-8pt}      %% reduce whitespace between graph and caption
\captionof{figure}{Light Contention} 
\label{Figure:graph-Light-Contention}
%% \end{figure}

\end{center} 
\vspace{5pt}

%% Actual name of benchmark was BU+BC but we use MutexBench for 
%% consistency with papers and as the name is more obvious. 

%% Discussion and interpretation of results

%% Figure \ref{Figure:graph-BU-64-1}, \ref{Figure:graph-BU-64-10} and \ref{Figure:graph-BU-64-10}
%% reflect uncontended locking performance collected with BU.  
%% As we drive up the number of threads and locks, performance fades for 
%% \texttt{hashChains} and \texttt{hashChains+F}, relative to \texttt{CJM} and \texttt{hashChains+3},
%% as collisions increase in the shared hash table.  In addition, in Figure \ref{Figure:graph-BU-64-1},
%% at one thread we see that \texttt{HashChains} provides relatively poor performance because of 
%% the latency required to acquire and release the bucket locks.  The same behavior can 
%% be seen at one thread in Figure \ref{Figure:graph-BC-1-1-0-0}.   
%% As noted above, under BU there is no true monitor contention.  

Figures \ref{Figure:graph-Maximum-Contention}, \ref{Figure:graph-Moderate-Contention} and
\ref{Figure:graph-Light-Contention} report the aggregate throughput of contended locking
operations with \texttt{MutexBench}. 
\texttt{synchronized}, \texttt{ReentrantLock} and \texttt{ReentrantLock-FIFO} 
reflect \texttt{MutexBench} ported to Java.  
\texttt{pthreads} is a degenerate version of our native C++ locking
framework where each object incorporates a \texttt{pthread\_mutex} lock.
\texttt{HashChain} and \texttt{HashChains+3} are described above.
\texttt{CJM-FIFO} is simple CJM and \texttt{CJM-By} is CJM with bounded bypass enabled.  

Figure \ref{Figure:graph-Maximum-Contention} shows extreme contention, with
empty critical and non-critical sections, on a single ``hot'' lock.  
$CSL$ and $NCSL$ are both configured as $0$.  
We see that \texttt{ReentrantLock-FIFO} scales poorly as the cost of park and unpark operations
and the futex operations are subsumed into the effective critical section length.
The other FIFO forms that use spin-then-block waiting (\texttt{CJM-FIFO, HashChains, 
and HashChains+3}) tend to cluster in a band.  
The spin duration is largely sufficient to avoid parking.
As noted above, \texttt{pthreads} fades 
because of futex sleep chain kernel spin lock contention induces by user-level
\texttt{pthread} contention.  \texttt{ReentrantLock} provides the best performance
closely followed by \texttt{synchronized}.  \texttt{ReentrantLock} and \texttt{synchronized} admit
long-term unfairness.  Over a 10 second interval, it is not uncommon to
see a 3x difference between the thread that accomplished the most iterations
versus the thread that completed the least iterations.   
\texttt{CJM-By} with bounded bypass performs almost as well as \texttt{synchronized} 
and \texttt{ReentrantLock} but is much fairer over the measurement interval. 

The data at 1 thread also serves as a good measure of uncontended latency.  
We can see, for instance, that \texttt{HashChains} exhibits increased latency,
because of the longer paths needed to acquire and release the bucket locks. 

%% Interestingly, in Figure \ref{Figure:graph-BC-1-1-0-0}, we see 
%% that the efforts in \texttt{HashChains+3} and \texttt{HashChains+F} to improve
%% latency, but also act against our interests and impinge on scalability. 

Figures \ref{Figure:graph-Moderate-Contention} ($CSL$=1 and $NCSL=200$) and 
\ref{Figure:graph-Light-Contention} ($CSL=1$ and $NCSL=1000$) 
also have a single hot lock, but use a very short critical section with longer non-critical sections,
reflecting more likely real-world scenarios.  Broadly, \texttt{CJM-By, synchronized} and
\texttt{ReentrantLock} provide comparable performance.  

CJM with bypass provides reasonable scaling -- in keeping and competitive with the existing 
\texttt{synchronized and ReentrantLock} implementation -- avoiding the performance collapses 
and retrograde scaling exhibited by some of the other locks.
In addition, CJM with bounded bypass provides long-term fairness, in contrast to
\texttt{pthreads}, \texttt{synchronized} and default \texttt{ReentrantLock}, but
at the same time remains preemption tolerant, unlike the FIFO locks.  

Not surprisingly, dedicating more header word state to synchronization yields
better synchronization performance.   The \texttt{HashChain} forms, which
are appealing simple, put little demand on the header, but suffer relative to 
the variants which use a displaced header word.  

It is worth remarking that our benchmarks used only a single object.  If we use 
multiple objects, then the \texttt{HashChain} forms start to suffer from 
false contention in the hash chains, even for logically uncontended locking.
%% where CJM avoids that pitfall.

\Invisible{Operating regimes; region of state space; realms}

%% \section{Variations} 

\section{Additional Remarks} 

\begin{enumerate}

\item The CJM variant should be able, with some additional effort, to tolerate GC algorithms that employ 
\emph{forwarding pointers}.  Presumably the low-order tag bits in the single header word
would encode the following possible states : \emph{Normal, Displaced-For-Locking, Forwarded}.  

\item If the JVM allows forwarding on-the-fly, by mutators, outside of safepoints, then an additional 
approach presents itself.  The first time we synchronize on a object, we immediately
copy and forward the object to an instance that has the CJM synchronization word appended 
to the object body.  At the same time, we change the object's type from $T$ to $T+$ to indicate
that the instance now supports the CJM word.  Conceptually we're cloning the object and
changing the type to a new derived class that contains the CJM word. 
For a given instance, the state transition is one way.   
Instead of relying on the type system, an additional bit in the header could also be 
reserved to indicate the \emph{has-lockword} state.
One concern inherent to this approach is that synchronization operations might
trigger out-of-memory conditions, although most synchronization designs also admit this same 
problem.

\item It is worth comparing the idea above to the ``tri-state'' approach reportedly used by some of IBM's JVM
implementations to reduce the size of the header.  2 bits are reserved in the header
to support the identity hashCode, encoding 3 possible states : \emph{neutral}, \emph{hashed-by-address},
and \emph{hashcode-appended}.  All objects start in neutral state.  On the 1st call to query
the identity hashCode, the JVM computes the hashCode as a deterministic function of the object's 
heap address, and shifts the state from neutral to hashed-by-address.   Subsequent calls 
to query the identity hashCode observe the hashed-by-address state and recompute and return
the same hashCode value.  If and when a hashed-by-address object is moved (copied) by the 
garbage collector, the collector changes the state to hashCode appended, and extends the 
object accordingly, computes the hashCode, and stores the identity hashCode value at the 
end of the object.  Subsequent calls to query the hashCode notice the hashcode-appended 
state and extract the value from the new field appended to the object.  
An unfortunate consequence of this tactic is that the collector can exhaust memory extending
objects during a garbage collection operation.  
See also \url{https://github.com/rkennke/lilliput/tree/compact-hashcode}.  

Related ideas can be found in IBM's J9 \emph{lock nursery} \cite{ecoop02-bacon} 
\url{https://blog.openj9.org/2019/04/02/lock-nursery/}

\item Instead of CJM, we could continue to use the existing
synchronization subsystem -- with all its inherent monitor lifecycle problems -- and displace 
the new lilliput single header into a compact \emph{displaced header word}, 
instead of the existing \emph{displaced mark} word.  
See \url{https://mail.openjdk.java.net/pipermail/lilliput-dev/2021-July/000096.html}. 
CJM, however, lends itself to displaced headers far more gracefully than the existing 
synchronized system.  

\item Both contended\cite{ols12} and uncontended performance are critical 
quality-of-implementation (\emph{QoI}) metrics for the design of a synchronization subsystem.  
Previously, in the era of \emph{bus locking}, before cache locking, techniques such
as \emph{biased locking}\cite{QRL} were used to improve the latency of uncontended operations. 
Thankfully, changes in processor architecture have obviated the need for biased locking,
and its incumbent complexity.  
Contended performance is usually measure in terms of scalability -- aggregate throughput and
Uncontended performance is measured via simple latency.  Between the extremes
we also find so-called ``promiscuous'' objects, which are locked by various threads but suffer
relatively little contention.  In this case, coherence traffic usually dictates performance.  
Specifically, an implementation should act to minimize write invalidation.  
While not as actively studied in the literature, performance is promiscuous mode is also of
importance. 

\item For the hash-based variants we'd want to employ a secondary hash function, likely
with a salted nonce, to reduce the threat of DoS attacks against the buckets. 

\item Interactions with projects \emph{loom} and \emph{graal} have not been considered.  

\item While not required, we assume to the extent reasonable and possible, that the object 
header will reside on the last word of a cache line, with the constituent object fields
starting on the following line.  This benefits SIMD and superword optimizations and
also reduces the span -- the number of cache lines -- underlying particular object instances,
improving spatial locality \footnote{A large fraction of current \texttt{malloc} allocators
get this precisely wrong, placing the header word on the first word of a cache line, even though
we expect the header word to be accessed infrequently relative to the words in the allocated block.}. 

\item We assume a 64-bit JVM with a 64-bit header word.   

\end{enumerate} 

%% \section{Raw Text}
%% 
%% \begin{enumerate}
%% 
%% \item Eliminate NL, NL, BU mentions
%% 
%% \item Use BC With T=1 for uncontended performance, perhaps plot with broken axes
%% 
%% \item JVM description; switches; etc
%% 
%% \item Java Methodology : 3x7 median
%% 
%% \item Bounded bypass with Fissile locks
%% 
%% \item ReentrantLock needs to allocate queue elements for each acquire, thus 
%% induces GC.  Sensitivity.  Sized heap to favor ReentrantLock. 
%% Also ABA avoidance
%% 
%% \item All hashchain forms described herein are FIFO but could add bounded bypass;  
%% Outer lock consists of reserved bit in object header word
%% 
%% 
%% \end{enumerate} 

\Invisible{Remote; displaced; away; xeno; off-object; decoupled; distal; distant} 

\bibliography{main}

\appendix

\section{Appendix -- Variations} 

%% HashChains+F form 
\BoldSection{HashChains with Fast-Path} : To address the issue of uncontended latency -- 
caused by the need to acquire and release the chain locks
for both acquire and release operations on the monitor --we can modify \textbf{HashChains} 
by adding a fast-path that allows threads to insert and remove a lock record from the bucket 
chain using just an atomic \texttt{compare-and-swap} (CAS) operation in the case where the 
chain is otherwise empty, avoiding the need to acquire and release the bucket locks.  We use a 
specialized encoding of the lock word for the bucket lock to indicate a singleton lock record is present.  
If additional threads (or objects) access the chain, the chain devolves to normal locking until 
the chain again becomes empty.  The fast-path optimization is purely a ploy to improve uncontended latency
\footnote{We collected performance data on this variant, but to avoid clutter in the graphs opted
to not report it below.}. 

%% CONSIDER : Subsection -- TACLL 
\BoldSection{CJM and Hashed Hybrid} : A hybrid of CJM and the hashed forms is also viable.  
Briefly, in this form,
we displace the header word as in CJM.  But to reduce complexity, we protect
the chain of waiting threads with internal locks instead of resorting to the lock-free techniques used in CJM.
This allows the chain to be structured as a tail-anchored circular linked list
(TACLL) which is convenient for our needs.  The header word points to the tail element 
(the most recently arrived) and the tail points to head, which is the owner.  
We also maintain the definitive displaced header word value in the tail element.  

\Invisible{Definitive, canonical, authoritative} 

The locks can be situated in either a shared global array hashed by the object's 
virtual address or hashCode, or we can site the locks in the chain elements.  
In the latter case, we lock the chain by attempting to lock the tail element and 
then confirming that the element remains at the tail, ``chasing'' ownership as needed.   
(This approach has lock-free progress properties.  If we find the tail changed then some other
thread made forward progress).  This variation requires safe memory reclamation
for the queue elements, but avoids hashing and hash collisions (false contention). 
The same locking protocol is used to access the displaced header word, which 
we maintain in the tail element referred to by the header word.  Inserting the initial
element, for uncontended acquisition, is accomplished with an atomic compare-and-swap operation
as there is not tail to lock.  This also accelerates uncontended operations.  

Access to the header word is accomplished as follows.  When the object 
is not locked the header is not displaced, and can thus be accessed directly.
Specifically, we can fetch the header word and examine the low order bits, which
constitute a discriminated union or tag, to determine if the object is locked.  
If the object is locked and the accessor is the owner (which is easily determined) then
the owner can simply find the displaced header word in the lock record it installed.
If the object is locked and the accessor is not the owner, a situation we expect
to be rare, then a more elaborate access protocol is required.  In this situation
the accessor locks the chain, as described above, to find the displaced header word 
in the tail element.  

This approach is simple and effective, but suffers from the ``double locking'' performance
concern.  In addition, both monitor acquire and release operations access the header word,
which is undesirable due to the increased coherence traffic on that location
\footnote{Again, we implemented and collected data on this variant, but opted to
omit the performance data in the evaluation section.}. 

\Invisible{chase/capture optimistic reading.} 
\Invisible{progress property is lock-free or obstruction-free but could be strengthened}

\end{document}